\documentclass[preprint,aps,amsmath,amssymb,twoside,prd,showkeys,superscriptaddress]{revtex4-1}
\pdfoutput=1
\usepackage{verbatim}
\usepackage{comment}
\usepackage{graphicx}
\usepackage[T1]{fontenc}
\usepackage{epsfig}
\usepackage{bm}
\usepackage{tensor}
\usepackage{amssymb}
\usepackage{float}
\usepackage{amsmath}
\usepackage{subfigure}
\usepackage{dcolumn}
\usepackage{cancel}
\usepackage[colorlinks]{hyperref}
\usepackage[usenames,dvipsnames]{color}
\hypersetup{
     breaklinks=true,
    pdfstartview={FitH},    
    colorlinks=true,       
    linkcolor=blue,          
    citecolor=red,        
    filecolor=magenta,      
    urlcolor=blue,           
    anchorcolor=green,      
    linktocpage=true
}

\def\doi{http://doi.org}

\newcommand{\HCd}{\mathcal{H}}

\def\HCdt0{\tilde{\HCd}_{0}}

\newcommand{\afffias}{Frankfurt Institute for Advanced Studies (FIAS), 
Ruth-Moufang-Strasse~1, 60438 Frankfurt am Main, Germany}
\newcommand{\affbgu}{Physics Department, Ben-Gurion University of the Negev, 
Beer-Sheva 
84105, Israel}
\newcommand{\affbahamas}{Bahamas Advanced Study Institute and Conferences, 4A 
Ocean 
Heights, Hill View Circle, Stella Maris, Long Island, The Bahamas}

\newcommand{\afftech}{The Technion Department of Physics,The Technion – Israel Institute of Technology,Haifa 3200003, Israel}

\newcommand{\affcam}{DAMTP, Centre for Mathematical Sciences, University of Cambridge, Wilberforce Road, Cambridge CB3 0WA, United Kingdom}

\begin{document}
\title{Noether Symmetry in Newtonian Dynamics and Cosmology}
\author{E. I. Guendelman}
\email{guendel@bgu.ac.il}
\affiliation{\affbgu}\affiliation{\afffias}\affiliation{\affbahamas}
\author{E. Zamlung}
\email{eyalzamlung@campus.technion.ac.il}
\affiliation{\afftech}
\author{D. Benisty}
\email{benidav@post.bgu.ac.il}
\affiliation{\affbgu}\affiliation{\afffias}\affiliation{\affcam}
\begin{abstract}
A new symmetry for Newtonian Dynamics is analyzed, this corresponds to going to an accelerated frame, which introduces a constant gravitational field into the system and subsequently. we consider the addition of a linear contribution to the gravitational potential $\phi$  which can be used to cancel the gravitational field induced by going to the accelerated from, the combination of these two operations produces then a symmetry. This symmetry leads then to a Noether current which is conserved. The conserved charges are analyzed in special cases. The charges may not be conserved if the Noether current produces flux at infinity, but such flux can be eliminated by going to the CM (center of mass) system in the case of an isolated system. In the CM frame the Noether charge vanishes,
Then we study connection between the Cosmological Principle and the Newtonian Dynamics which was formulated via a symmetry \cite{Benisty:2019wpm} of this type, but without an action formulation. Homogeneous behavior for the coordinate system relevant to cosmology leads to a zero Noether current and the requirement of the Newtonian potential to be invariant under the symmetry in this case yields the Friedmann equations, which appear as a consistency condition for the symmetry.

\end{abstract}
\maketitle
\section{Introduction}
In this paper we will discuss a  new basic
symmetry on the Newton's Second Law (NSL) and of the Poisson equation which allows to introduce a uniform acceleration in space, but not
constant in time. In order to obtain a symmetry of the Newtonian equations of motion and Poisson equation,  the Newtonian potential is
also transformed accordingly. In general relativity a uniformly accelerating reference frame and a frame at rest in uniform gravitational field are nonequivalent as dicussed by Deslodge\cite{EdwardADesloge}, see also related discussion by Singleton et. al. \cite{Jones:2007qy},
, however if we consider a transformation in position which will give as a linear contribution to the gravitational potential
it will be correct for the newtonian case, where this complication does not exist, where there is a unique definition of a uniform gravitational field. Two transformations are involved in the symmetry: One performs the transformation to an accelerated system, and the other one adds a linear contribution to the gravitational potential. The transformation in position to an accelerated frame according to the equivalence principle of Einstein produces a uniform gravitational field and the addition of the linear contribution produces also a constant gravitational field that can exactly cancel the one generated by going to the accelerated frame. We obtain therefore a symmetry. Such a symmetry was discussed at the level of the equations of motion and applied to cosmology \cite{Benisty:2019wpm}, which allows considerations of homogeneity of Newtonian cosmology. From the uniform expansion of the universe one can correlate going from one comoving observer to another to going to an accelerated frame, the consistenct of the symmetry is shown to gives rise to the Friedmann equations. 

For the complete discussion of the symmetry and conserved quantities, one must introduce an action principle and show that the above symmetries are indeed symmetries of the action. Here we show that a simple mechanical system of point particles that incorporates a potential $\phi$, yields a symmetry which according to Noether \cite{noether1971invariant} leads to conserved quantities.
From Noether theorem, any symmetry gives a conservation law. This allows us to calculate the conserved Noether symmetry.
The conserved charges are analyzed in special cases. 
In general the conservation law involves a charge density and a current of such charge and we obtain a local conservation and the the conservation is obtained automatically if the current does not have a flux at infinity, 
The charges may not be conserved however if the Noether current produces flux at infinity, but such flux can be eliminated by going to the CM system in the case of an isolated system. In the CM frame the Noether charge vanishes,
In a relativistic theory the charge density and the current form together a 4 vector, while in a non-relativistic case this is not so, but still there will be a charge density, a current and a local conservation law. 
We then go back to the cosmological case, which was analized before without the use of Noether currents \cite{Benisty:2019wpm} .
Cosmology without having General Relativity (GR) can be formulated from Newton’s theory \cite{Liddle:1998ew,Sanghai:2017grt,Concha:2019dqs,Hanimeli:2019svd,Barrow:2020wmk,Casadio:2020nns,Castro-Palacio:2020ssz,ellis2013discrete,ellis2015discrete}  and beyond cosmology Newtonian gravity, together with a geometric interpretation of the results,  can provide a way to obtain a grasp on gravitatational collapse and the Schwarzschild solution \cite{Guendelman:2018ndb}. The Friedmann equation that describes the expansion of the universe has a form of kinetic part $\dot{a}^2$ and a potential-like part $V(a)$ \cite{Benisty:2019vej}, where $a$ is the scale factor of the universe. \cite{Benisty:2019wpm} derives the Friedmann equations from a symmetry. In particular, while one frame introduces a homogeneous expansion, where Hubble law is valid with respect to any point in the universe, \cite{Benisty:2019wpm} shows that the homogeneity of the universe emerges from those two effects that cancel each other and yield the Friedmann equations. From the point of view of the Noether charges found in this paper, we see that these symmetries consisting of going from one comoving observer to another leads to a conserved quantity that for the homogeneously expanding universe is identically zero. This represents a vacuum that respects the symmetry, i.e., there is no spontaneous symmetry breaking in this situation, This resembles the result obtained for the vanishing of the conserved charge in the CM of an isolated system, if one thinks that any point in a uniform universe can be though of as the CM of the universe. Finally from the condition of invariance of the Newtonian potential under the symmetry in cosmology, we obtain the Friedmann equations.

\section{The Mechanical System}
If the sources of the gravitational field are point masses, we must add to the Lagrangian of the gravitational field a term which is a function of  potential and the coordinates of the ith particle \cite{anderson1967principles}.
In the General Theory of relativity, gravity, described by a metric and a system of point particles are described by the action,
\begin{eqnarray}
S&=&S_g+S_{m} \label{SgSphiSm}
\label{totaction}\\
 S_g&=&-\frac{1}{\kappa}\int \sqrt{-g}R(\Gamma,g) d^{4}x \,,
\nonumber\\
S_{m}&=&\int \sqrt{-g}L_m d^{4}x \, , \nonumber
\nonumber
\end{eqnarray}
where
\begin{eqnarray}
R(\Gamma,g)=g^{\mu\nu}\left( \Gamma^{\lambda}_{\mu\nu
,\lambda}-\Gamma^{\lambda}_{\mu\lambda ,\nu}+
\Gamma^{\lambda}_{\alpha\lambda}\Gamma^{\alpha}_{\mu\nu}-
\Gamma^{\lambda}_{\alpha\nu}\Gamma^{\alpha}_{\mu\lambda}\right)
\nonumber
\end{eqnarray}
and the Lagrangian for the matter, as collection of particles $S_m$ reads
\begin{equation}
L_m=-\sum_{i} m_i \int 
\sqrt{g_{\alpha\beta}\frac{dx_i^{\alpha}}{d\lambda}\frac{dx_i^{\beta}}{d\lambda}}\,
\frac{\delta^4(x-x_i(\lambda))}{\sqrt{-g}}d\lambda \label{Lm}
\end{equation}
where $\lambda$ is an arbitrary parameter. We have the freedom to choose $\lambda$ as we wish. A convenient and standard choice is to choose all the particle times equal and equal to this parameter $\lambda$, then the integration over this  $ t$ gives rise to a three dimensional delta function.

Here we will be concerned with the much simpler non-relativistic approximation of such theory, which is represented by Newtonian gravity interacting with a system of point particles. The role of the metric is taken by the Newtonian potential And our work will be within the framework of Noether's theorem in classical field theories \cite{FieldNoetherTheorem}. The starting point is then the following Lagrangian:
\begin{equation}
L = \sum\limits_i {\left( {\frac{1}{2}{m_i}{{\dot {\vec x}_i}}^2 - {m_i}\phi({\vec x}_i) } \right) - \frac{1}{{8\pi G}}\int {{{\left( {\vec \nabla \phi } \right)}^2}} {d^3}x}   
\label{eq:lag}
\end{equation}
that describes a system of $i$ particles. $m_i$ is the mass of the ith point particle, $G$ is the universal Newtonian Constant, $x(t)_i$ is the location vector of the ith point particle, and $\phi(x)$ is the potential. Equivalently, the Lagrangian density can be constructed that is the analog of the one that we have described above in the GR case:
\begin{equation}
{\cal L} =\sum\limits_i {\left( {\frac{1}{2}{m_i}{{\dot {\vec x}_i}}^2 - {m_i}\phi } \right){\delta ^3}\left( {\vec x - {{\vec x}_i}(t)} \right) - \frac{1}{{8\pi G}}{{\left( {\vec \nabla \phi } \right)}^2}}  
\label{eq:lagden}
\end{equation}
Notice of course that $\delta ^3 \left( {\vec x - {{\vec x}_i}(t)} \right)\phi(\vec x)= \delta ^3\left( {\vec x - {{\vec x}_i}(t)} \right)\phi({\vec x}_i (t)) $ . 
The corresponding definition for the Lagrangian in terms of the Lagrangian density: 
\begin{equation}
L = \int \mathcal{L} \, d^3 x
\end{equation}
and the action being given by
\begin{equation}
S = \int L dt 
\end{equation}
The symmetry include two transformations as we will discuss. The transformation for the coordinates, which represents going to another frame which has an arbitrary 
acceleration with respect to the original one reads:
\begin{equation}
\vec x' = \vec x + \varepsilon {\mkern 1mu} {\vec x_0}(t).
\label{eq:corTr}
\end{equation}
Here $ {\vec x_0}(t)$ is a three dimensional vector with an arbitrary dependence on time which describes the tranformation from one frame to another which is accelerated with respect to it.
The second transformation shifts the gravitational potential in order to introduce a constant gravitational field that cancel the acceleration. Notice that:
\begin{equation}
\phi '({\vec x^\prime }) = \phi (\vec x) - \varepsilon {\ddot {\vec x}_0} \cdot \vec x.   
\end{equation}

In principle it is possible to add an arbitrary function of time to the Newtonian potential as well, and then we will have the generalized form

\begin{equation}\label{aceTra}
\ddot{x}_{i}'(t)-\ddot{x_{i}}(t)=g_{i}(t)
\end{equation}
\begin{equation}\label{potTra}
\phi'(x_{j}',t)=\phi[x_{i}(x_{j}'),t]-g_{k}(t)x_{k}(x_{j}')+h(t)
\end{equation}
where $\forall i,j,k=1,2,3.$. $h(t)$ is an arbitrary function of time that can be added to the Newtonian potential and $g_{i}(t)$ denoting the relative acceleration between the two systems. 
For the moment however we will set $h(t)= 0$, although for some purposes it may be useful to consider $h(t) \neq 0$, see for example in our discussion of homogeneity at the end of the paper. Transformations of this type were considered long time ago also by Schucking  \cite{Shucking1967cosmology}, but not as a Lagrangian symmetry, only as a symmetry of the equations of motion. The symmetry of the equations of motion under such transformations are very simple to prove. First of all the gravitational field introduced by the relative acceleration is cancelled by the opposite gravitational field induced by the linear term in the coordinates which is also propostional to the coordinates. Second, a linear term in the coordinates in the Newtonian potential leaves the Poisson equation unchanged , since the secont derivative of the linear term being added is zero. 
Because of the shift of the coordinates (Eq. \ref{eq:corTr}) in addition the shift of the potential the final expression for the transformation gives:
\begin{equation}
\begin{split}
\phi '(\vec x') = \phi '(\vec x) + \varepsilon {\vec x_0} \cdot \vec \nabla \phi ' = \phi '(\vec x) + \varepsilon {\vec x_0} \cdot \vec \nabla \phi  \\ = \phi (\vec x) - \varepsilon {\ddot {\vec x}_0} \cdot \vec x + O({\varepsilon ^2}).
\end{split}
\end{equation}
In simplified form one can formulate the transformation as:
\begin{equation}
\phi '(\vec x) = \phi (\vec x) - \varepsilon {\ddot {\vec x}_0} \cdot \vec x - \varepsilon {\vec x_0} \cdot \vec \nabla \phi .
\end{equation}
Extending the transformation into all space besides the particle coordinate, yielding the total transformation for the gravitational potential:
\begin{equation}
\phi '(\vec x) = \phi (\vec x) - \varepsilon \left( {{{\ddot {\vec x}}_0} \cdot \vec x + {{\vec x}_0} \cdot \vec \nabla \phi } \right).
\end{equation}
The Lagrangian has three parts, the kinetic part of the particle, the kinetic part of the quadratic gradient part of the potential and the interaction part. The transformation on the Lagrangian yields a total divergence. The transformation of the kinetic term yields:
\begin{equation}
\begin{split}
 \delta {{\cal L}_{kin}}/\varepsilon =  \sum\limits_i  {m_i}{{\dot {\vec x}}_i} \cdot {{\dot {\vec x}}_0}{\delta ^3}(\vec x - {{\vec x}_i}(t)) \\ - \frac{1}{2}{m_i}{{\dot{\vec x}}_i^2}{\mkern 1mu} {{\vec x}_0} \cdot \vec \nabla {\delta ^3}(\vec x - {{\vec x}_i}(t))  .  
\end{split}
\end{equation}
The transformation on the interaction reads:
\begin{equation}
\begin{split}
\delta {{\cal L}_{int}}= \varepsilon \sum\limits_i ( {  {m_i}\phi (\vec x){{\vec x}_0} \cdot \vec \nabla {\delta ^3}(\vec x - {{\vec x}_i}(t))  +{m_i}{{\vec x}_0} \cdot \vec \nabla \phi {\delta ^3}(\vec x - {{\vec x}_i}(t)) + {m_i}{{\ddot {\vec x}}_0} \cdot {{\vec x}_i}{\delta ^3}(\vec x - {{\vec x}_i}(t))}) .    
\end{split}
\end{equation}
The transformation on the quadratic gradient of the gravitational potential gives:
\begin{equation}
 \delta {{\cal L}_{quad}} = \varepsilon \left( {\frac{{\vec \nabla \phi  \cdot \vec \nabla ({{\vec x}_0} \cdot \vec \nabla \phi ) + {{\ddot {\vec x}}_0} \cdot \vec \nabla \phi }}{{4\pi G}}} \right).
\end{equation}
The transformation of the kinetic term $\delta\mathcal{L}_{kin}$ contains the contribution of the form:
\begin{equation}
\begin{split}
 {m_i}{\dot {\vec x}_i} \cdot {\dot {\vec x}_0}{\delta ^3}\\ = {m_i}\frac{\partial }{{\partial t}}({\vec x_i} \cdot {\dot {\vec x}_0}{\delta ^3}(\vec x - {\vec x_i}(t))) - {m_i}{\vec x_i} \cdot {\ddot {\vec x}_0}{\delta ^3}(\vec x - {\vec x_i}(t))) \\ -{m_i}{\vec x_i} \cdot {\dot {\vec x}_0}\frac{\partial }{{\partial t}}{\delta ^3}(\vec x - {\vec x_i}(t)).   
\end{split}
\label{eq:term13}
\end{equation}
The last term can be express as special derivatives using chain rules:
\begin{equation}
- {m_i}{\vec x_i} \cdot {\dot {\vec x}_0}\frac{\partial }{{\partial t}}{\delta ^3}(\vec x - {\vec x_i}(t))) = {m_i}{\dot {\vec x}_i}(t) \cdot \vec \nabla {\delta ^3}(\vec x - {\vec x_i}){\vec x_i} \cdot {\dot {\vec x}_0}.
\end{equation}
Setting this in Eq. (\ref{eq:term13}) while taking in account the chain rule for the Delta function:
\begin{equation}
\begin{split}
 {m_i}{\dot {\vec x}_i} \cdot {\dot {\vec x}_0}{\delta ^3}(\vec x - {\vec x_i}(t)) = \frac{\partial }{{\partial t}}({m_i}{\vec x_i} \cdot {\dot {\vec x}_0}{\delta ^3}(\vec x - {\vec x_i}(t))) + \nabla ({m_i}{\dot {\vec x}_i}(t){\delta ^3}(\vec x - {\vec x_i}(t)){\vec x_i} \cdot {\dot \vec x_0}) \\
 - {m_i}{\vec x_i} \cdot {\ddot {\vec x}_0}{\delta ^3}(\vec x - {\vec x_i}(t)) - \nabla \left( {{m_i}{{\dot \vec x}_i}({{\vec x}_i}(t) \cdot {{\dot {\vec x}}_0})} \right){\delta ^3}(\vec x - {\vec x_i}(t))   
 \end{split}
\end{equation}
Because the last term vanishes, we get:
\begin{equation}
\begin{split}
  {m_i}{\dot {\vec x}_i} \cdot {\dot {\vec x}_0}{\delta ^3}(\vec x - {\vec x_i}(t)) = \frac{\partial }{{\partial t}}({m_i}{\vec x_i} \cdot {\dot {\vec x}_0}{\delta ^3}(\vec x - {\vec x_i}(t))) \\+ \nabla ({m_i}{\dot {\vec x}_i}(t){\delta ^3}(\vec x - {\vec x_i}){\vec x_i} \cdot {\dot {\vec x}_0}) - {m_i}{\vec x_i} \cdot {\ddot {\vec x}_0}{\delta ^3}(\vec x - {\vec x_i}(t))     
\end{split}
\end{equation}
We take the gradient out side. Therefore the complete transformation reads:
\begin{equation}
\begin{split}
\delta \mathcal{L}/\varepsilon  = \frac{\partial }{{\partial t}}\left( {\sum\nolimits_i {{m_i}{{\vec x}_i} \cdot {{\dot {\vec x}}_0}(t){\delta ^3}(\vec x - {{\vec x}_i}(t))} } \right) \\
+ \nabla \left( {\sum\nolimits_i {\left( {{m_i}{{\dot {\vec x}}_i}\left( {{{\dot {\vec x}}_0} \cdot {{\vec x}_i}} \right) - \frac{1}{2}{m_i}{{\dot {\vec x}}_i}^2{{\vec x}_0} + {m_i}\phi {{\vec x}_0}} \right)} {\delta ^3}(\vec x - {{\vec x}_i}(t))} \right)
\\  + \nabla \cdot \left( {\frac{{{{\ddot {\vec x}}_0}\phi }}{{4\pi G}}} \right) + \frac{{\vec \nabla \phi  \cdot (\vec \nabla ({{\vec x}_0}(t) \cdot \vec \nabla \phi )}}{{4\pi G}}
\end{split}
\end{equation}
The ${\raise0.7ex\hbox{${\delta {{\cal L}_{quad}}}$} \!\mathord{\left/
 {\vphantom {{\delta {{\cal L}_{quad}}} \varepsilon }}\right.\kern-\nulldelimiterspace}
\!\lower0.7ex\hbox{$\varepsilon $}}$ can be shown to be a total derivative. The term $ \vec \nabla \phi  \cdot \vec \nabla ({\vec x_0} \cdot \vec \nabla \phi )$ can be written in two different ways. The first one:
\begin{equation}
=  \nabla (\vec \nabla \phi  \cdot \vec \nabla ({\vec x_0} \cdot \vec \nabla \phi )) - \nabla^2 \phi \vec{x}_0 \cdot \vec\nabla \phi
\label{eq:term1}
\end{equation}
emerges from the cancellation of $\nabla^2 \phi \vec{x}_0 \cdot \vec\nabla \phi$. The second way:
\begin{equation}
=   \nabla (\phi \vec \nabla  \cdot ({\vec x_0} \cdot \vec \nabla \phi )) - \phi {\nabla ^2}({\vec x_0} \cdot \vec \nabla \phi )
\label{eq:term2}
\end{equation}
emerges from the cancellation of $\phi \nabla^2 (\vec{x}_0 \cdot \vec \nabla\phi)$. The average of Eq. (\ref{eq:term1}) and Eq. (\ref{eq:term2}) gives:
\begin{equation}
 = \frac{1}{2}\nabla  \cdot \left[ {  \vec \nabla ({{\vec x}_0} \cdot \vec \nabla \phi ) + \phi \vec \nabla ({{\vec x}_0} \cdot \vec \nabla \phi ) - {{\vec x}_0}\phi {\nabla ^2}\phi } \right].
\end{equation}
The complete transformation reads:
\begin{equation}
\begin{split}
 \delta \mathcal{L}/\varepsilon  = \frac{\partial }{{\partial t}}\left( {\sum\nolimits_i {{m_i}{{\vec x}_i} \cdot {{\dot {\vec x}}_0}(t){\delta ^3}(\vec x - {{\vec x}_i}(t))} } \right) \\
 + \nabla \left( {\sum\nolimits_i {\left( {{m_i}{{\dot {\vec x}}_i}\left( {{{\dot {\vec x}}_0} \cdot {{\vec x}_i}} \right) - \frac{1}{2}{m_i}{{\dot {\vec x}}_i}^2{{\vec x}_0} + {m_i}\phi {{\vec x}_0}} \right)} {\delta ^3}(\vec x - {{\vec x}_i}(t))} \right)
 \\
  + \nabla \left( {\frac{{{{\ddot {\vec x}}_0}\phi }}{{4\pi G}} + \frac{{\vec \nabla (\phi {{\vec x}_0} \cdot \vec \nabla \phi ) - {{\vec x}_0}\phi {\nabla ^2}\phi }}{{8\pi G}}} \right)  
 \end{split}
\end{equation}
Since the integral of a divergence becomes a boundary term according to the divergence theorem, the action is still invariant under the corresponding symmetry. Moreover, a conserved NC can be obtained.

The transformation of the Lagrangian is a total derivative of the vector: 
\begin{equation}
\Lambda^{0} = \sum_i m_i \vec{x}_i\cdot \dot{\vec{x}}_0 \delta^3 (\vec{x} - \vec{x}_i(t))
\end{equation}
\begin{equation}
\begin{split}
\vec{\Lambda} =  {\sum\limits_i {\left( {{m_i}{{\dot {\vec x}}_i}\left( {{{\dot {\vec x}}_0} \cdot {{\vec x}_i}} \right) - \frac{1}{2}{m_i}{{\dot {\vec x}}_i}^2{{\vec x}_0} + {m_i}\phi {{\vec x}_0}} \right)} {\delta ^3}(\vec x - {{\vec x}_i}(t))}  \\
+ {\frac{{{{\ddot {\vec x}}_0}\phi }}{{4\pi G}} + \frac{{\vec \nabla (\phi {{\vec x}_0} \cdot \vec \nabla \phi ) - {{\vec x}_0}\phi {\nabla ^2}\phi }}{{8\pi G}}}
\end{split}
\end{equation}
where:
\begin{equation}
\Lambda^{\mu} = \left(\Lambda^{0},\vec{\Lambda}\right).
\end{equation}
Consequently, the total divergence reads:
\begin{equation}
{\raise0.7ex\hbox{${\delta {\cal L}}$} \!\mathord{\left/
 {\vphantom {{\delta {\cal L}} \varepsilon }}\right.\kern-\nulldelimiterspace}
\!\lower0.7ex\hbox{$\varepsilon $}} = {\partial _\mu }{\Lambda ^\mu }
\end{equation}
The corresponding NC \cite{noether1971invariant} yields:
\begin{equation}
j^{\mu} = (j^{0}, j^{i}),
\end{equation}
where:
\begin{equation}
{j^0} = {\Lambda ^0} - \frac{{\partial {\cal L}}}{{\partial {{\dot x}_i}}} \cdot {\raise0.7ex\hbox{${\delta {{\dot x}_i}}$} \!\mathord{\left/
 {\vphantom {{\delta {{\dot x}_i}} \varepsilon }}\right.\kern-\nulldelimiterspace}
\!\lower0.7ex\hbox{$\varepsilon $}},
\end{equation}
\begin{equation}
{j^i} = {\Lambda ^i} - \frac{{\partial {\cal L}}}{{\partial ({\partial _i}\phi )}} \cdot {\raise0.7ex\hbox{${\delta \phi }$} \!\mathord{\left/
 {\vphantom {{\delta
 \phi } \varepsilon }}\right.\kern-\nulldelimiterspace}
\!\lower0.7ex\hbox{$\varepsilon $}}.
\end{equation}
Notice that the current involves the coordinate and the potential. The variation gives:
\begin{equation}
{j^0} = \sum\limits_i {{m_i}} ({\vec x_i} \cdot {\dot {\vec x}_0} - {\dot {\vec x}_i} \cdot {\vec x_0}){\delta ^3}(\vec x - {\vec x_i}(t))
\label{eq:grstate}
\end{equation}
\begin{equation}
\begin{split}
\vec j = {\sum\limits_i {\left( {{m_i}{{\dot {\vec x}}_i}\left( {{{\dot {\vec x}}_0} \cdot {{\vec x}_i}} \right) - \frac{1}{2}{m_i}{{\dot {\vec x}}_i}^2{{\vec x}_0} + {m_i}\phi {{\vec x}_0}} \right)} {\delta ^3}(\vec x - {{\vec x}_i}(t))}  
\\
 + {\frac{{{{\ddot {\vec x}}_0}\phi }}{{4\pi G}} + \frac{{\vec \nabla (\phi {{\vec x}_0} \cdot \vec \nabla \phi ) - {{\vec x}_0}\phi {\nabla ^2}\phi }}{{8\pi G}}}
 - \frac{{\vec \nabla \phi }}{{4\pi G}}\left( {{{\ddot {\vec x}}_0}\cdot \vec x + {{\vec x}_0}\cdot \vec \nabla \phi } \right)
\end{split}
\end{equation}
One can study the conserved charge $Q  = \int d^3x j^0  =
\sum\limits_i {{m_i}} ({\vec x_i} \cdot {\dot {\vec x}_0} - {\dot {\vec x}_i} \cdot {\vec x_0}) $ for a couple of special cases;

\begin{enumerate}
    \item $ {\vec x}_0 = constant $, which gives $Q=-{\vec x}_0\cdot \textbf{P} $, where $\textbf{P}= 
\sum\limits_i {{m_i}}  {\dot {\vec x}_i} $ is the total momentum of the system of particles interacting through gravitational interactions.
Since  $ {\vec x}_0 $ is an arbitrary constant vector that can point in any direction, we obtain in this case the conservation of the linear momentum as a consequence of the translation invariance of the theory. 
\item  $ {\vec x}_0 =  {\vec v}_0 t$, which gives  $Q({\vec v}_0)= (M\textbf{X}_{CM}- \textbf{P}t)\cdot {\vec v}_0 $, which implies that the center of mass moves with a constant velocity.

The general expression for  the charge in terms of the center of mass position and the total momentum of the mutually interacting set of particles is,

$Q= (M\textbf{X}_{CM} \cdot {\dot {\vec x}}_0- \textbf{P}\cdot {\vec x}_0) $.

Notice that when ${\ddot{\vec x}}_0 \neq 0$ , there is a longer range contributions to the current $\vec j $ of the form 
$- \frac{{\vec \nabla \phi }}{{4\pi G}}\left( {{{\ddot {\vec x}}_0}\cdot \vec x  } \right)$ and $\frac{{{{\ddot {\vec x}}_0}\phi }}{{4\pi G}}$ which could potentially produce a non vanishing flux of the charge at infinity and therefore a potential non conservation of such a charge. In this respect it is also worth point out that in the Center of Mass frame $\textbf{P}=0 $ and $\textbf{X}_{CM}=0$ and then $Q=0$ for any $ {\vec x}_0 $.  When choosing the origin of coordinates at the CM of the system, assuming a multiple  expantion for the Newtonian potential, we can check that the overall flux of $- \frac{{\vec \nabla \phi }}{{4\pi G}}\left( {{{\ddot {\vec x}}_0}\cdot \vec x  } \right)$ and $\frac{{{{\ddot {\vec x}}_0}\phi }}{{4\pi G}}$ becomes zero for the monopole contribution, because of cancellation of positive and negative contributions and zero for the dipole contribution, because the dipole is zero in the CM frame. The additional multipoles do not produce flux at infinity  and therefore  the conserved quantity is recovered, even for  ${\ddot{\vec x}}_0 \neq 0$ and this conserved charge is then zero in the CM frame.
In this case one is insured of a conserved charge for an arbitrary choice of  $ {\vec x}_0 $, in fact for a generic choice, which is beyond the constant 
or linear in time $ {\vec x}_0 $  one must choose the CM frame to get the conserved quantity, which is then zero, or the frame that is obtained from the CM frame through the accelerated transformation. 

Indeed, notice also that in the transformed frame from the CM frame through the accelerated transformation, $\textbf{X}_{CM}= {\vec x}_0$ and $\textbf{P}=M{\dot{\vec {x}}_0} $, which also lead us to

$Q= (M\textbf{X}_{CM} \cdot {\dot {\vec x}}_0- \textbf{P}\cdot {\vec x}_0) $.
$={M\vec x}_0 \cdot {\dot {\vec x}}_0 - M\dot{\vec x}_0\cdot {\vec x}_0 = 0$.

Zero again, although the transformed frame is not anymore the CM frame.
The vanishing of the conserved quantity is also a consequence of the fact that when we demand local symmetries (here meaning the symmetry involving arbitrary functions of time), the only consistent associated charges are zero, for example, the application of the Noether Theorem to local gauge invariance in Electrodynamics leads to the conserved quantity $\vec \nabla\cdot \vec E - 4\pi \rho $, which is of course identically zero.
\end{enumerate}

For time dependent $G$ the symmetry is preserved, since the symmetry does not depend on the $G$, and the integration by parts that are needed in the last term of the action involving the newton constant are only spacial. Therefore homogeneous cosmology is guaranteed  also for these theories, which have been formulated in  \cite{Hanimeli:2019svd}.   

It is important to point out that there are alternative ways to investigate Newtonian gravity that may shed additional light on the symmetries study here. In particular, Newton-Cartan geometry \cite{Ruede:1996sy,Hansen:2020pqs} which is based on Galilean symmetries which, has
also uncovered connections to cosmology. It seems therefore that it will be useful to put the results in this broader context and study the possibility of studying these symmetries in the formalism provided by such approaches.

\section{The origin of Hubble’s law from the homogeneity of the universe and why transforming from one point to another corresponds to transforming to an accelerated frame }
Each comoving observer in an homogeneously expanding universe expands with a diferent velocity, and we still want to claim that the universe is homogeneous, how is this possible?. The solution is simple, it is just that the symmetry that allows us is a symmetry that allows us to express this homogeneity, i.e., that allows us  to transform from one point to another corresponds to transforming to an accelerated frame.
If we denote the coordinate of the original co-moving observer as unprimed whereas the coordinate of another object, which is also a comoving 
 object but moves with respect to the initial object as primed. Then, the relative velocity of the object with
respect to the original observer can be written as $\vec v' = \vec v'\left( {\vec r',t} \right)$ .  Now, if we take our origin to a general
point and if the relative displacement vector between the original observer and the primed object is ${\vec s}$ then the displacement vector and velocity vector becomes: $\vec r' = \vec r - \vec s$ and $\vec v'\left( {\vec r - \vec s,t} \right)$ . Using the law of transformation of velocities between two accelerated frames,  we conclude that:
\begin{equation}
\vec v'\left( {\vec r - \vec s,t} \right) = \vec v\left( {\vec r,t} \right) - \vec v\left( {\vec s,t} \right)
\end{equation}
Therefore the velocity vector must be linear in the position vector.This implies that if we want a velocity dependence on the position vector that is reproduced with respect to any comoving point in the expanding universe, we must have indeed that the velocity must be  linear in the displacement vector, that is $v_{i} =f_{ij}r_{j}$. If we add isotropy to our considerations, this leads us with the only choice, which is to conceive $f_{ij}= f  \delta_{ij} $ , or $\vec v\left( {\vec r,t} \right) = f  \vec r$,    where $f$ is a function independent of ${\vec r}$  which is indeed Hubble´s law.
It must be dependent solely on time, so that $f = f\left( t \right)$.
So we can write the velocity vector in the following way:
\begin{equation}
\vec v\left( {\vec r,t} \right) = f\left( t \right)  \vec r
\label{eq:velocity_vector}
\end{equation}
The most important thing is that this relation holds with respect to any point in the homogeneous universe, so the Hubble law is perfectly consistent with the homogeneity of the universe, although at first sight the Hubble law seems to prefer a special point as the center of the universe,  this is not the case, once velocities are approprietly transformed,  the velocities in the frame where a certain point is considered at rest, then other points  have velocities linear with distance with respect to such point, and Hubble´s law is recovered with respect to any comoving observer in the universe. This is easily seen, since
\begin{equation}
\vec v'\left( {\vec r - \vec s,t} \right) = \vec v\left( {\vec r,t} \right) - \vec v\left( {\vec s,t} \right) =  f\left( t \right) (\vec r - \vec s)
\end{equation}
After integration,  equation (\ref{eq:velocity_vector}) gives rise to:
\begin{equation}
\vec r = R\left( t \right) \cdot {{\vec r}_0}
\end{equation}
where, $R\left( t \right)$ satisfies
\begin{equation}
\frac{1}{R}\frac{{dR}}{{dt}} = f\left( t \right),R\left( {t = 0} \right) = 1
\end{equation}
This equation is an expression of the Hubble’s constant in terms of the scale factor $H\left( t \right) = f\left( t \right)$. $R\left( t \right)$ is
nothing but a scale factor representing the homogeneous expansion of the universe.

\section{Symmetry transformations and vanishing conserved quantities in the Homogeneous Cosmological Solution}
As explained in the previous section, we can consider a case where all particle coordinates are expanding according to the same expansion factor:
\begin{equation}
\vec x \equiv \vec \chi  R\left( t \right)
\label{eq:expansion}
\end{equation}
where $\chi$ is a constant "comoving coordinate" and $R(t)$ is the scale factor of the universe. The shift of the comoving coordinate by a constant, which , as we demonstrated in the previous section  allows us to transfer one comoving point of the homogeneous universe to another comoving point, is of course associated with a transformation to an accelerated frame:
\begin{equation}
\vec \chi ' \equiv \vec \chi  + \varepsilon \vec c
\end{equation}
from where we can identify the vector ${\vec x_0}$ that describes the transformation to the accelerated frame to be:
\begin{equation}
{\vec x_0}\left( t \right) = \vec c  R\left( t \right)
\label{eq:shift}
\end{equation}
We can see that under this symmetry, we obtain that the associated Noether charge density identically vanishes, since 
now from Eq. (\ref{eq:grstate}) get the form of:
\begin{equation}
\sum\limits_i {{m_i}\left( {{{\vec \chi }_i} \cdot \vec cR\dot R - {{\vec \chi }_i} \cdot \vec c\dot RR} \right){\delta ^3}(\vec x - {{\vec x}_i}(t)) = 0}  
\end{equation}

This means that the symmetry is unbroken, since this vacuum has zero charge.
This resembles the result obtained in the previous section that
the conserved quantity is zero in the CM, so may be another way of thinking about this result is thinking that any point in an homogeneous universe can be think of its CM.
For this unbroken vacuum we also expect an invariant Newtonian potential, which would mean that the shift of the coordinate cancel the transformation of the gravitational field, so we require that:
\begin{equation}
{{\ddot {\vec x}}_0} \cdot \vec x + {{\vec x}_0} \cdot \vec \nabla \phi=0
\end{equation}
by inserting Eq. (\ref{eq:expansion}) and Eq. (\ref{eq:shift}) to the last equation, we get:
\begin{equation}
\vec \chi \ddot R\left( t \right) \cdot \vec c R\left( t \right) + \vec c R\left( t \right) \cdot \vec \nabla \phi  = 0
\end{equation}
taking into account the solution to poisson equation for a constant density ($\phi \left( {\vec x} \right) = {4\pi G\rho }/3\cdot {\vec x^2}$) and inserting it in the above equation, we receive:
\begin{equation}
\vec \chi  \cdot \vec c\ddot R + \frac{{4\pi G\rho }}{3}R\vec c \cdot \vec \chi  = 0
\label{eq:Rfactor}
\end{equation}
for Eq. (\ref{eq:Rfactor}) to be true for any ${\vec c}$
, we conclude:
\begin{equation}
\frac{{\ddot R}}{R} =  - \frac{{4\pi G\rho }}{3}
\end{equation}
which is also known as the second order Friedmann equation (for pressureless dust era), which appears here as a consequence of the requirement of invariance of this vacuum describing homogeneous expansion of the universe. The density here represents a universe with matter. In order to formulate the cosmological constant $\Lambda$, the potential is modified to the term:
\begin{equation}
\mathcal{L}_\Lambda = \frac{1}{32\pi G} \phi  \Lambda
\end{equation}
In such a way, the contribution for the Poisson equation gets the uniform contribution that is known from the effect of $\Lambda$. 

\section{Cosmological Solution from shift symmetry}
So far we have been studying infinitesimal transformations, which is all we need as far as the conserved quantities are concerned. We want to see however 
how the symmetry looks like for the finite case. We will see that it is convenient for this purpose to write 
\begin{equation}
\label{comExp}
 x_i = \chi_i R(t)   
\end{equation}
where $R(t)$ is the time dependent scale factor of the universe. Now we check the validity of the translation invariance of the space:
\begin{equation}\label{trans}
\chi_i \rightarrow \chi_i + c_i,
\end{equation}
where $c_i $ denotes a constant vector. Therefore the position transforms according to the relation: 
\begin{equation}
x_i \rightarrow x_i + c_i R(t) = x'_i,
\end{equation}
and the local acceleration transforms as:
\begin{equation}
\ddot{x}_i \rightarrow \ddot{x}_i + c_i \ddot{R}(t).
\end{equation}
From Eq. (\ref{aceTra}) we identify the relative acceleration $g_i(t)$ as:
\begin{equation}\label{gi}
g_i(t) = c_i \ddot{R}(t).
\end{equation}
the isotropic solution for the Poisson equation reads: 
\begin{equation}\label{tranPoten}
\phi=\frac{2\pi G\rho (t)}{3}x_{i}x_{i}.
\end{equation}
Due to the uniform background, the coordinates should be valid considering any cartesian coordinate system centered around any point.
Notice that our assumption that the density $\rho$ is a function of time comes from the cosmological principle. It is now our task to show that despite appearances, Eq. (\ref{tranPoten}) doesn't single out a special point in the universe, since the transformations (\ref{aceTra})-(\ref{potTra}) imply for cosmology that all points are on an equal footing. As we will see then,  the potential will be well defined for any arbitrary origin.
The transformation of the gravitational potential, from Eq. (\ref{tranPoten}), reads:
\begin{equation}\label{tranPotenS}
\begin{split}
\phi'(x')=\frac{2\pi G\rho}{3}x'_{i}x'_{i} = \frac{2\pi G\rho}{3} (x_i + c_i R(t))(x_i + c_i R(t)) \\ = \frac{2\pi G\rho}{3} x_i x_i + 
\frac{4\pi G\rho}{3} x_i c_i R(t) +  \frac{2\pi G\rho}{3}c_i c_i R(t)^2.   
\end{split}
\end{equation}

This transformation corresponds to Eq. (\ref{potTra}) only for:
\begin{equation}\label{gii}
g_i(t) = -\frac{4\pi G\rho}{3} c_i R(t) , \quad h(t) = \frac{2\pi G\rho}{3}c_i c_i R(t)^2 
\end{equation}
Notice that in the infinitesimal case h can be ignored since it is quadratic in c, but it appears in the finite transformation.
For a consistency between Eq. (\ref{gi}) Eq. (\ref{gii}) and Eq. (\ref{aceTra}), we get:
\begin{equation}
c_i \ddot{R}(t)=-\frac{4\pi G\rho}{3} c_i R(t),
\end{equation}
which reduces to the relation:
\begin{equation}\label{Friedmann_eq}
\frac{\ddot{R}(t)}{R(t)} = -\frac{4\pi G \rho}{3}.
\end{equation}
So, the Friedmann equations appear as a consequence of the homogeneity of the universe.
For deriving a second equation, we first consider mass conservation within co-moving sphere \cite{Guendelman:2018ndb},
\begin{equation}
    \frac{d}{{dt}}\left( {\frac{{4\pi }}{3}{R^3}\rho } \right) = 0
\end{equation}
where the internal mass inside the sphere should be conserved. By performing the derivative
and simplifying one R, the equation gets:
\begin{equation}
    2R\dot R\rho  + {R^2}\dot \rho  + R\dot R\rho  = 0
\end{equation}
The last term can be eliminated by Eq. (\ref{Friedmann_eq}), and after restoring derivatives the equation
\begin{equation}
    \frac{{d\left( {{{\dot R}^2}} \right)}}{{dt}} = \frac{{8\pi G}}{3}\frac{{d\left( {{R^2}\rho } \right)}}{{dt}}
\end{equation}
is obtained. Integration on both sides gives
\begin{equation}
    {\dot R^2} = \frac{{8\pi G}}{3}{R^2}\rho  - \tilde k
\end{equation}
and rewriting the arbitrary integration constant $\tilde k$ in a way to match the units $\tilde k \to k{c^2}$ yields
finally an equation, which corresponds to Friedmann's first equation:
\begin{equation}
    {\left( {\frac{{\dot R}}{R}} \right)^2} = \frac{{8\pi G}}{3}\rho  - k{\left( {\frac{c}{R}} \right)^2}.
\end{equation}
 If we think about it, it is very much like a conserved energy of a mechanical system and indeed the
sign of k determines whether R will expand to infinity, or expand, achieve a maximum and then
re collapse. This can be compared with a mechanical system in a following explanatory way:
For k > 0, we have a bound system, a bound system that has an associated negative energy
For k < 0, represents a case where the expansion factor reaches infinity like that of a mechanical
system with positive energy.
For k = 0, produces the case where R barely makes it to infinity and corresponds to a system with
zero energy, analogous to the very similar problem of the critical escape trajectory that just makes
in order to escape the earth.
\section{Discussions and Conclusions}
First of all we have here studied a new symmetry present in the case of a system of particles interacting through the gravitational dynamics of Newton.
The Equivalence principle shows that we can introduce a global uniform acceleration all over space which has the same effect of introducing a uniform 
gravitational field. A uniform gravitational field can also be induced by adding a linear term in the coordinates, this linear addition to the gravitational potential can be chosen so as to exactly cancel the gravitational field induced by going to the uniformly accelerated gravitational field, obtaining in this case a symmetry by the combination of these two transformations.  
In a Lagrangian formulation a system of classical particles governed by a gravitational field contains  Noether currents and conserved quantities  that follow from these symmetries, since the general continuous transformation connects two different frames with relative acceleration and the shifting of gravitational potential (in order to cancel the acceleration) leave the Lagrangian density invariant up to a total derivative. 

The conserved quantities are displayed for some particular cases, leading to the well known conservation of momentum, for a simple translation, the conservation of the velocity of the center of mass, in the case of a a transformation linear in time. We then see that the conserved charge is identically zero in the CM of a localized system, in this frame the potentially dangerous asymptotic behavior of the currents at infinity that may invalidate the conservation of the charge dissapears. 

 We finally see the correspondence between this symmetry and the homogeneity of the universe. We examine the application of these transformations in Newtonian Cosmology and see that they appear naturally when we consider translation of "co-moving coordinates". We show then that under the Newtonian gravity, translation of co-moving coordinates in a uniformly expanding universe defines a new accelerated frame. The Friedmann equations appear as a consistency condition  for the symmetry.

\acknowledgments
EG thanks for Ben Gurion University of the Negev for  support. DB thanks for Ben Gurion University of the Negev and Frankfurt Institute for Advanced Studies for  great support. This article is supported by COST Action CA15117 "Cosmology and Astrophysics Network for Theoretical Advances and Training Action" (CANTATA) of the COST (European Cooperation in Science and Technology). This project is partially supported by COST Actions CA16104 and CA18108.

\bibliographystyle{apsrev4-1}
\bibliography{Bibliography}

\end{document}